\input{aipcheck}
\documentclass[final]{aipproc}
\layoutstyle{6x9}
\begin{document}
\centerline{}
\vspace{-2.2truecm}
\title{Extracting infrared QCD coupling from meson spectrum}
\classification{12.38.Aw, 11.10.St, 12.38.Lg, 12.39.Ki}
%\classification{<Replace this text with PACS numbers; choose from this list:
%                \texttt{http://www.aip..org/pacs/index.html}>}
\keywords      {Running coupling, QCD, meson spectrum}
\author{M.~Baldicchi, G.~M.~Prosperi, C.~Simolo}{address={Dipartimento di
Fisica, Universit\`a di Milano\\
I.N.F.N., sezione di Milano\\
via Celoria 16, I20133 Milano, Italy}}

%******************************************************************************
%******************************************************************************

%\documentclass[12pt]{article}
%\usepackage{graphicx}
%\usepackage[dvips]{color}
%\newcommand{\be}{\begin{equation}}
%\newcommand{\ee}{\end{equation}}
%\newcommand{\bea}{\begin{eqnarray}}
%\newcommand{\eea}{\end{eqnarray}}
%\newcommand{\nn}{\nonumber}
%\newcommand{\lb}{\label}

%\adresss{Dipartimento di Fisica dell'Universit\'a di Milano\\
%Istituto Nazionale di Fisica Nucleare,Sezione di Milano\\
%Via Celoria 16,I20133 Milano, Italy\\E-mail: prosperi@mi.infn.it}

%******************************************************************************
%******************************************************************************

\begin{abstract}
In the framework of the Bethe-Salpeter formalism used in previous 
papers to evaluate the quarkonium spectrum, here we reverse the
point of view to extract an ``experimental'' running coupling 
$\alpha_{\rm s}^{\rm exp}(Q^2)$ in the infrared (IR) 
region from the data. The values so obtained agree within 
the erros with the Shirkov-Solovtsov analytic coupling for 
200 MeV $<Q<$ 1.2 GeV, thus giving a very satisfactory unifying description of
high and low energy phenomena. Below 1~GeV however $\alpha_{\rm s}^{\rm exp}(Q^2)$
seems to vanish as $Q\to0\,$. The paper is based on a work in progress 
in collaboration with D.~V.~Shirkov. 
\end{abstract}

\maketitle
%000000000000000000000000000000000000000000000000000000000
\vspace{-0.5truecm}
%The present paper is based on a work in progress in collaboration  
%with D.~V.~Shirkov.

As well known a very consistent picture of the high
energy processes can be obtained by perturbative QCD if the
running coupling $\alpha_{\rm s} ( Q^{2} )\,$,
as derived from the renormalization group, is used 
and a good convergence is already attained at 3-loop level 
(see e.g.~\cite{pdg}). In the traditional $\overline {\rm MS}$ 
renormalization scheme, however, $\alpha_{\rm s} ( Q^{2} )$ 
develops at any loop level unphysical singularities for 
$Q\sim\Lambda_{\rm QCD}$ (Landau singularities) that make the 
expression useless in the IR region. This 
is a serious difficulty in any quark model where $Q$ should be 
identified with the momentum transfer taking values typically 
between few GeV and some hundred MeV. 

Among the various attempts to eliminate Landau singularities 
(see e.g. \cite{Prosperi:2006hx}) we consider  
the proposal of Shirkov and Solovtsov, which consists in imposing 
analyticity on $\alpha_{\rm s} ( Q^{2})\,$\cite{shirkov}. 
At 1-loop the analytic coupling can be written explicitly
\begin{equation}
  \alpha_{\rm an}^{(1)} ( Q^{2} ) = \frac{1}{
   \beta_{0} } \left(
  \frac{1}{ \ln{ ( Q^{2} / \Lambda^{2} ) } } +
  \frac{ \Lambda^{2} }{ \Lambda^{2} - Q^{2} } \right)\,.
\label{eq:1loop}
\end{equation}
At 2- or 3-loop level $\alpha_{\rm an} ( Q^{2} )$ can be only 
numerically computed. At 3-loop an useful approximation is however 
given by the ``1-loop-like'' model
\begin{equation}
 \alpha_{\rm an}^{(3)} ( Q^{2} ) = \frac{ 4 \pi }{
   \beta_{0} } \left(
  \frac{1}{l} +
  \frac{1}{1 - e^l} \right), \quad {\rm with}\quad
  l = \ln{ Q^{2} \over \Lambda^{2} } + {\beta_1 \over \beta_0^2} \ln
\sqrt{\ln^2{ Q^{2} \over \Lambda^{2} } + 2\pi^2}
\label{eq:3loop}
\end{equation}
and the proper $\overline{\rm MS}$ value for $\Lambda\,$. 
For $500\;{\rm MeV}<Q<200 \;{\rm GeV}$ eq.\ (\ref{eq:3loop}) differs 
from the exact expression by no more than 2\%. Furthermore below 1~GeV 
$\alpha_{\rm an}^{(3)} ( Q^{2})$ with $\Lambda_{n_f=3}^{(3)}=375$~MeV 
differs even less from eq.\ (\ref{eq:1loop}) with  
$\Lambda_{n_f=3}^{(1)}=206$~MeV. 

On the other side in the last years we have developed a Bethe
-Salpether formalism like \cite{BMP} that was applied with a certain
success to the
calculation of the meson spectrum in the light and
in the heavy quark sectors. The formalism was essentially derived from
QCD first principles, making only an ansatz on the Wilson loop correlator 
$W\,$, which consists in writing $i\ln W$ as the sum of a one-gluon exchange 
and an area term encoding confinement, 
$i\ln W = (i\ln W)_{\rm OGE}+ \sigma S \,$.
The resulting reduced Salpeter equation is then in the form of the eigenvalue
equation for a squared bound state mass 
\begin{equation}
    M^2 = M_0^2 + U_{\rm OGE}+U_{\rm CF} \,,
\label{eq:M2}
\end{equation}
where $ M_0 = w_1+w_2 = \sqrt{m_{P1}^2 + 
{\bf k}^2} + \sqrt{m_{P2}^2 + {\bf k}^2}$ and $U=U_{\rm OGE}+U_{\rm CF}$ 
the potential (see \cite{BMP, BP} and references 
therein). By neglecting the spin orbit 
and the the tensorial term but including the hyperfine splitting term, 
$ U_{\rm OGE}$ has the form
\begin{equation}
\langle {\bf k} \vert U_{\rm OGE}  \vert {\bf k}^\prime \rangle=
    \rho\,{4\over 3} {\alpha_{\rm s}({\bf Q}^2) \over \pi^2}\,
\bigg[ - \frac{1}{{\bf Q}^2}
     \bigg( q_{10} q_{20} + {\bf q}^2 - { ( {\bf Q}\cdot {\bf q})^2 \over
{\bf Q}^2 } \bigg) + {1\over 6} {\bf \sigma}_1 \cdot {\bf \sigma}_2 
    \bigg ]\quad
\label{upt}
\end{equation}
where $\rho$ is a kinematic factor. In \cite{BP} 
we have computed the meson masses by the equation $ m^2_{a}
=\langle\phi_a|M_0^2|\phi_a\rangle+\langle\phi_a|U_{\rm OGE}|\phi_a\rangle+
\langle\phi_a|U_{\rm CF}|\phi_a\rangle\,$, where $\phi_a$ is  
the zero-order wave function for the state $a$ obtained by solving the 
eigenvalue equation for the static limit Hamiltonian 
$H_{\rm CM} = w_1 + w_2  - {4 \over 3} {\alpha_{\rm s} \over r } +  \sigma
r \,$ by the Rayleigh-Ritz method. To this a second order correction in 
the hyperfine term was added in some cases.
 
\begin{figure}%[htbp!]
\begin{picture}(420,150)
 \put(20,0){\includegraphics[height=.25\textheight,
                      width=.32\textheight]{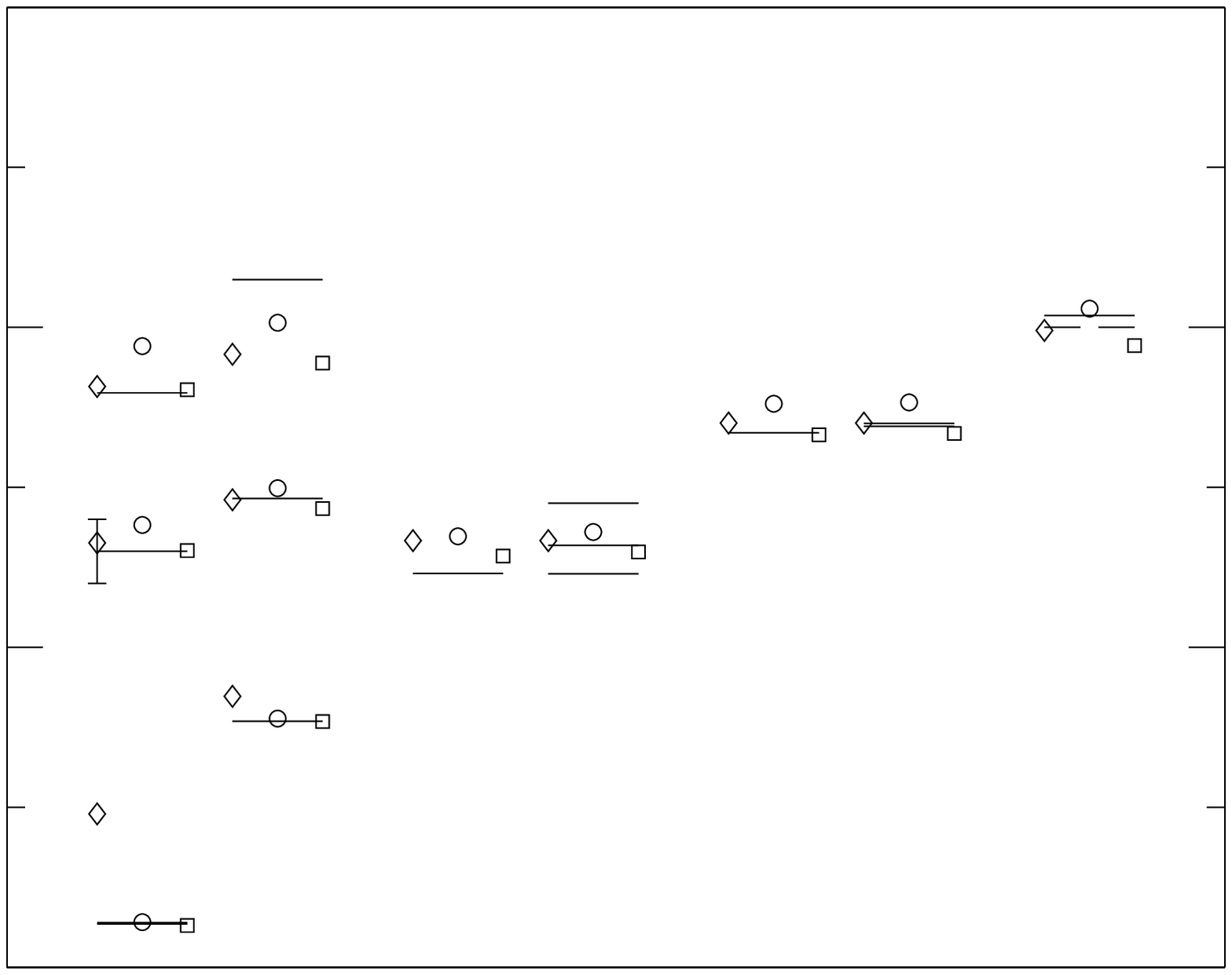}}
 \put(235,0){\includegraphics[height=.25\textheight,
                       width=.32\textheight]{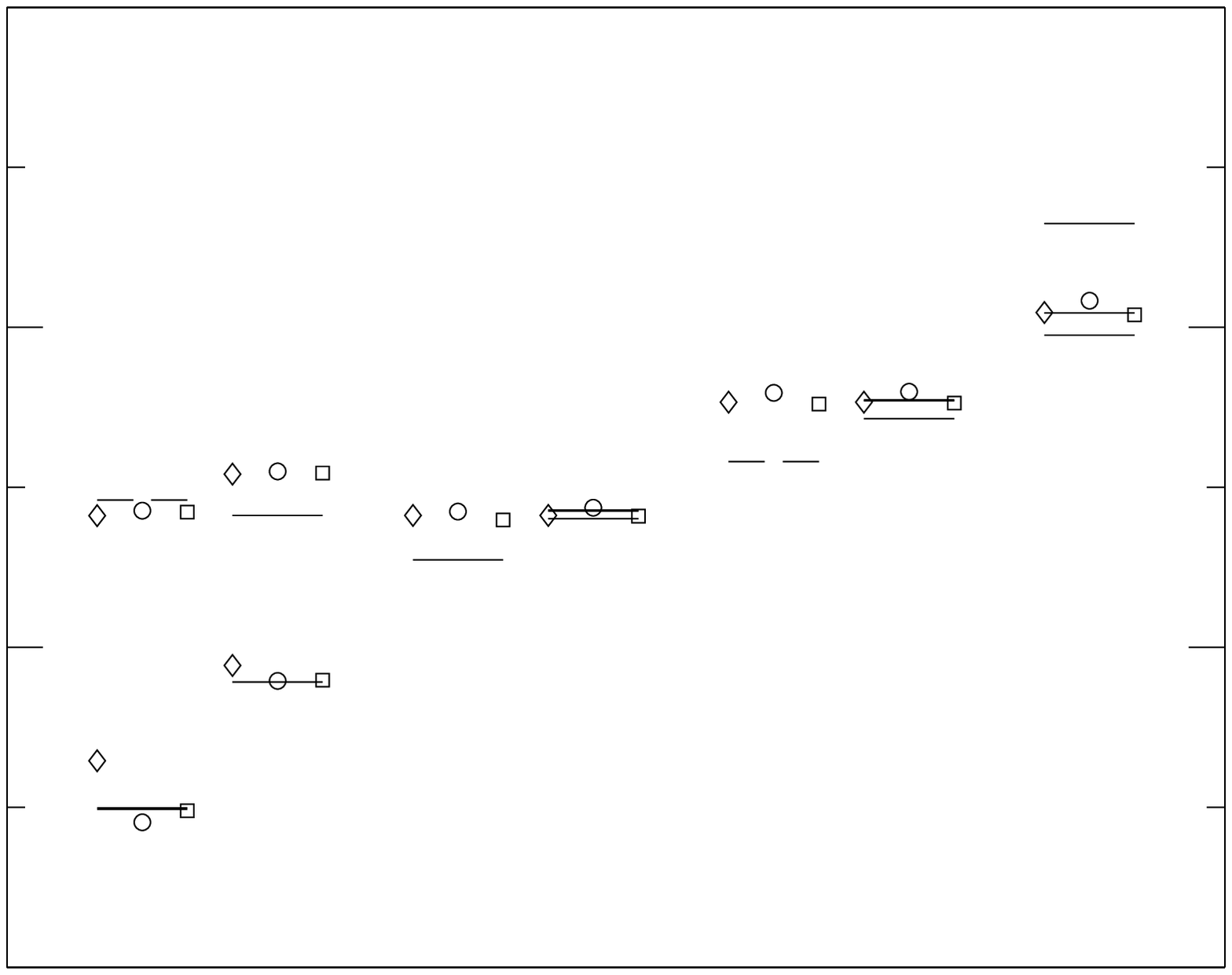}}
  \put(5,165){ $ {}_{ ( {\rm MeV} ) } $ }
       \put(0,150){$ {}_{3000} $}
       \put(0,103){$ {}_{2000} $}
       \put(0,55){$ {}_{1000} $}
       \put(13,2){$ {}_{0} $}
       \put(183,25){ $ q \bar{q} $ }
       \put(153,10){ $ ( q = u,d ) $ }
       \put(28,135){ $ {^{1} {\rm S}_{0}} $ }
       \put(51,135){ $ {^{3} {\rm S}_{1}} $ }
       \put(78,135){ $ {^{1} {\rm P}_{1}} $ }
       \put(100,135){ $ {^{3} {\rm P}_{J}} $ }
       \put(128,135){ $ {^{1} {\rm D}_{1}} $ }
       \put(150,135){ $ {^{3} {\rm D}_{J}} $ }
       \put(180,135){ $ {^{3} {\rm F}_{J}} $ }
       \put(55,5){ $ \pi $ }
       \put(75,36){ $ \rho $ }
%%%%      
      %\put(216,437){$ {}_{3000} $}
      %\put(216,392){$ {}_{2000} $}
      %\put(216,347){$ {}_{1000} $}
      %\put(229,302){$ {}_{0} $}

       \put(400,25){ $ q \bar{s} $ }
      \put(370,10){ $ ( q = u,d ) $ }
       \put(243,135){ $ {^{1} {\rm S}_{0}} $ }
       \put(266,135){ $ {^{3} {\rm S}_{1}} $ }
       \put(294,135){ $ {^{1} {\rm P}_{1}} $ }
       \put(317,135){ $ {^{3} {\rm P}_{J}} $ }
       \put(344,135){ $ {^{1} {\rm D}_{1}} $ }
       \put(367,135){ $ {^{3} {\rm D}_{J}} $ }
       \put(397,135){ $ {^{3} {\rm F}_{J}} $ }
       \put(270,22){ $ K $ }
       \put(292,43){ $ K^{\ast} $ }
    \end{picture}
\vspace{-3.5truecm}
\caption{\footnotesize Quarkonium spectrum, three different calculations.  
Diamonds refer to the truncation prescription for $\alpha_{\rm s}\,$, 
squares and circles refer to the 1-loop analytic coupling (\ref{eq:1loop}) 
and two different parametrizations for constituent masses of light quarks. 
Lines represent experimental data.
\vspace{-0.5truecm}}
\label{spect}
\end{figure}

Calculations have been performed by using both a truncation prescription 
for $\alpha_{\rm s}(Q^2)$ and the 1-loop analytic
coupling (\ref{eq:1loop}). The results of three sets of calculations are
graphically reported in Fig.~\ref{spect} for the light-light and 
light-strange sectors as an example. The key point is that, while the two
different assumptions on $\alpha_{\rm s}(Q^2)$ give similar results for
the heavy-heavy quark states a correct reprodution 
of the $\pi$ and $K$ masses can be obtained, as it can be seen, only with 
the analytic coupling. 

\begin{figure}[!ht]
\includegraphics[width=0.85\textwidth, height=0.35\textheight, angle=-360]{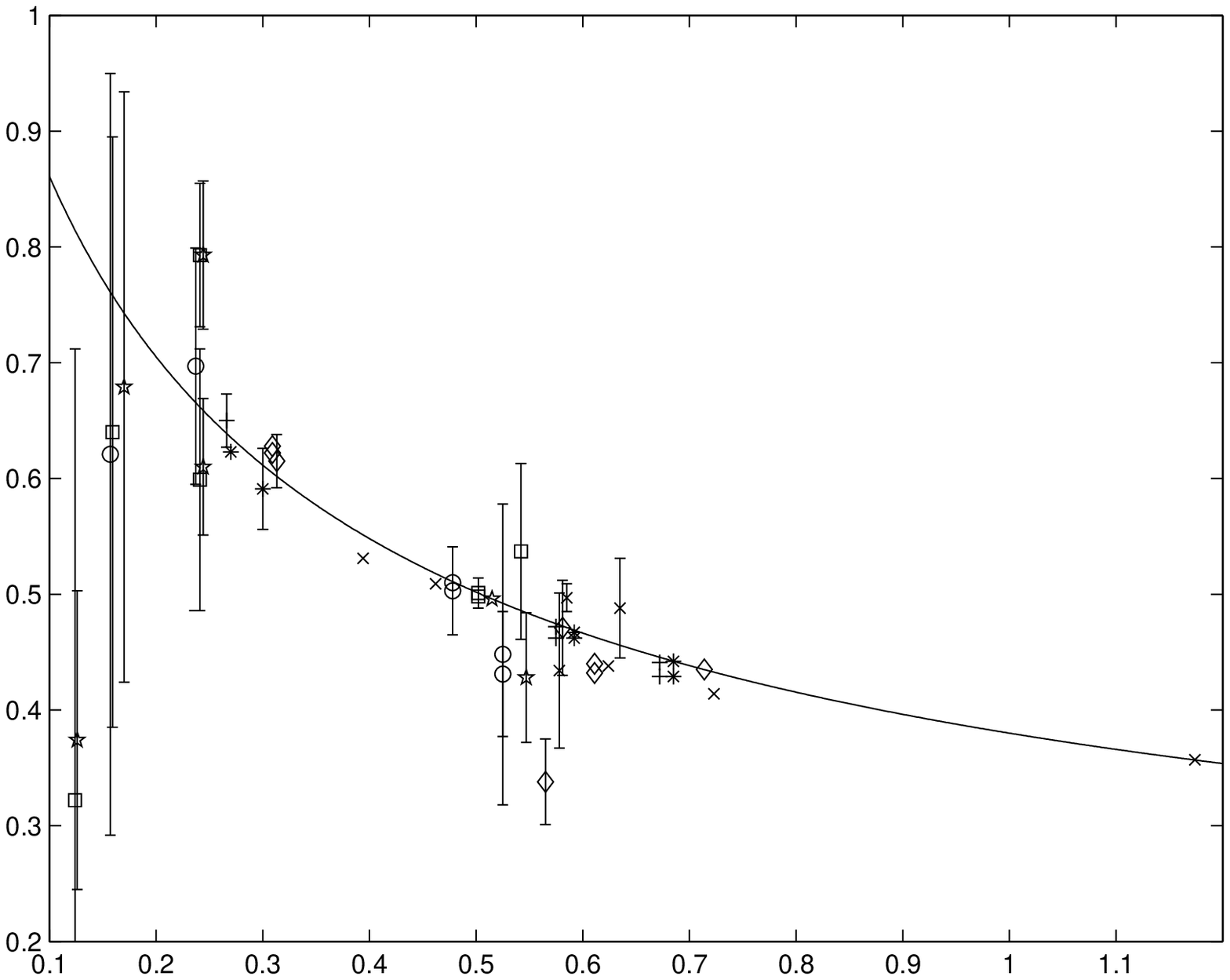}
\put(-290,190){ {\footnotesize $\alpha_{\rm s}(Q^2)$} }
\put(-70,25){ {\footnotesize $Q\,\,$[GeV]} }
\put(-100,190){ {\footnotesize $\sigma=0.18\,\,{\rm GeV}^2$} }
\put(-100,170){ {\footnotesize $\Lambda_{n_f=3}^{(1)}=206$~MeV} }
\put(-100,150){ {\footnotesize $m_u=m_d=211$~MeV} }
\put(-100,130){ {\footnotesize $m_s=306$~MeV} }
\put(-100,110){ {\footnotesize $m_c=1.524$~GeV} }
\put(-100,90){ {\footnotesize $m_b=4.865$~GeV} }
%{\color{red}{
%\put(-316,113){\vector(1,2){15.5}}
%\put(-324,104){{\footnotesize $ \rho $}}
%\put(-316,185){\vector(1,-2){15.3}}
%\put(-327,186){{\footnotesize $ \pi $}}
%\put(-237,205){\vector(-1,-4){17}}
%\put(-240,211){{\footnotesize $ \varphi $}}
%\put(-180,148){\vector(-1,-1){26}}
%\put(-180,150){{\footnotesize $ J/ \psi $}}
%\put(-63,116){\vector(1,-1){26}}
%\put(-76,116){{\footnotesize $ \Upsilon $}}
%}}
\caption{{\footnotesize 3-loop analytic coupling with 
$\Lambda^{(3)}_{n_f=3}=375$~MeV and  $\alpha_{\rm s}^{\rm exp}$. 
Circles, stars and squares refer respectively to $q\bar q\,$, 
$s\bar s$ and $q\bar s$ with $q=u,d\,$; diamonds and crosses stay 
for $c\bar c$ and $b\bar b\,$, plus signes for $q\bar c$ and $q\bar b$, 
while asterisks for $s\bar c$ and $s\bar b\,$. Error bars are drawn only 
if relevant.}} 
\label{figX}
\vspace{-0.3truecm}
\end{figure}

In this paper we focus our attention on the reversed point of view. 
$\Lambda_{n_f=3}^{(1)}$ and quark masses have been fixed by fitting 
$\pi\,$, $\rho\,$, $\phi\,$, $J/\psi\,$ and $\Upsilon$ mesons, while 
the string tension has been fixed a priori to the value 
$\sigma = 0.18\,\,{\rm GeV}^2\,$ (see Fig.~\ref{figX}). 
For each state $a$ we then define a theoretical fixed coupling 
$\alpha_{{\rm s}, a}^{\rm th}$ which leads to the same theoretical mass
as by using $\alpha_{\rm an}^{(1)}(Q^2)\,$. Thus an effective momentum 
transfer $Q_a$ is assigned to each state by the equation 
$\alpha_{\rm an}^{(1)}(Q_a^2)=\,\alpha_{{\rm s}, a}^{\rm th}\,$.
We finally define $\alpha_{\rm s}^{\rm exp}(Q_a^2)$ as the 
value of the coupling to be inserted in (\ref{upt}) in order to exactly 
reproduce the experimental mass:  
$\langle\phi_a|M_0^2|\phi_a\rangle+
\alpha_{\rm s}^{\rm exp}(Q_a^2)\langle\phi_a|{\mathcal O}({\bf q};
{\bf Q})|\phi_a\rangle +
\langle\phi_a|U_{\rm CF}|\phi_a\rangle=m^2_{\rm exp}\,$ 
(${\mathcal O}({\bf q};{\bf Q})$ given by (\ref{upt})). 
\vspace{-0.2truecm}

The results are given pictorially on Fig.~\ref{figX}; points representing 
$\alpha_{\rm s}^{\rm exp}(Q_a^2)$ are compared with the analytic curve
(\ref{eq:3loop}) for $\Lambda_{n_f=3}^{(3)}=375$~MeV. 
Error bars take into account the theoretical errors in the determination 
of the spectrum as well as the experimental ones when relevant. 
The theoretical incertitude expected in our procedure, that does not 
include coupling among different channels, is assumed to be roughly 
expressed by the half width of the state. As it can been seen the 
$\alpha_{\rm s}^{\rm exp}(Q_a^2)$ values agree rather well with the 
analytic coupling expression within the quoted errors for 
$200\,{\rm MeV}< Q< 1.2 \,{\rm GeV}$. Below 200 MeV, however, there 
seems to exist a consitent tendency of $\alpha_{\rm s}^{\rm exp}(Q_a^2)$ 
to vanish rather than to approach a finite limit.\\
The present paper is based on a work in collaboration with D.~V.~Shirkov.

\end{document}